\documentclass[10pt,twocolumn]{IEEEtran}

\usepackage{amsmath,amssymb,euscript,yfonts,psfrag,latexsym,dsfont,graphicx,bbm,color,amstext,wasysym,pdfsync,soul,framed,url,balance,soul}
\renewcommand{\baselinestretch}{1}
\graphicspath{{/},{figures/}}

\newcommand{\E}{{\mathbb E}}

\newcommand{\cL}{{\mathcal L}}

\newcommand{\fP}{{P}}

\newcommand{\fM}{{\mathfrak M}}

\newcommand{\D}{{\mathbb D}}

\newcommand{\qed}{\hfill $\Box$ \vskip 2ex}

\newcommand{\var}{\operatorname{\rm Var}}
\newcommand{\diag}{\mathop{\mathrm{diag}}}

\newcommand{\y}{{x_0}}

\newcommand{\support}{{\rm Supp}}

\definecolor{llgrey}{rgb}{0.9,0.9,0.9}
\definecolor{lgrey}{rgb}{0.6,0.6,0.6}
\definecolor{lred}{rgb}{0.9,0.7,0.7}

\newtheorem{theorem}{Theorem}

\newtheorem{prop}{Proposition}
\newtheorem{problem}{Problem}
\newtheorem{definition}{Definition}
\newtheorem{remark}{Remark}

\def \proof {\noindent {\it Proof. }}

\IEEEoverridecommandlockouts

\def\spacingset#1{\def\baselinestretch{#1}\small\normalsize}
\spacingset{.9}

\begin{document}

\title{\huge {Efficient-robust routing for single commodity network flows}\thanks{This project was supported by AFOSR grants (FA9550-15-1-0045 and FA9550-17-1-0435), grants from the National Center for Research Resources (P41-RR-013218) and the National Institute of Biomedical Imaging and Bioengineering (P41-EB-015902), National Science Foundation (ECCS-1509387), by the University of Padova Research Project CPDA 140897 and a postdoctoral fellowship through Memorial Sloan Kettering Cancer Center.
}}
\author{Yongxin Chen, Tryphon T. Georgiou, {\em Fellow IEEE}, Michele Pavon, and Allen Tannenbaum, {\em Fellow IEEE}
\thanks{Y.\ Chen is with the Department of Medical Physics, Memorial Sloan Kettering Cancer Center, NY; email: chen2468@umn.edu}
\thanks{T.\ T. Georgiou is with the Department of Mechanical and Aerospace Engineering, University of California, Irvine, CA; email: tryphon@uci.edu}
\thanks{M. Pavon is with the Dipartimento di Matematica ``Tullio Levi Civita",
Universit\`a di Padova, 35121 Padova, Italy; email: pavon@math.unipd.it}
\thanks{A.\ Tannenbaum is with the Departments of Computer Science and Applied Mathematics \& Statistics, Stony Brook University, NY; email: allen.tannenbaum@stonybrook.edu}}


\maketitle

\spacingset{1}
\begin{abstract}
 We {study single commodity network flows with suitable robustness and efficiency specs. An original use of a maximum entropy problem for distributions on the paths of the graph turns this problem into a {\em steering problem} for Markov chains with prescribed initial and final marginals. }
From a computational standpoint, viewing scheduling this way
is especially attractive in light of  the existence of an iterative algorithm to compute the solution. 
The present paper builds on \cite{chen2016networks} by introducing an  index of {\em efficiency} of a transportation plan and points, accordingly, to efficient-robust transport policies. In developing the theory, we establish two  new {\em invariance properties} of the solution (called {\em bridge}) 
-- an {\em iterated bridge} invariance property and the invariance of the {\em most probable paths}. These properties, which were tangentially mentioned in our previous work, 
are fully developed here. We also show that the distribution on paths of the optimal transport policy, which depends on a ``{\em temperature}'' parameter, 
tends to the solution of the ``most economical'' but possibly less robust {\em optimal mass transport} problem as the temperature goes to zero.  The relevance of all of these properties for transport over networks is illustrated in an example.
\end{abstract}

\spacingset{.9}
 
\spacingset{.97}
\section{Introduction}
{Consider a company owning a factory $F$ and a warehouse $W$. The company wants to ship a certain quantity of goods from $F$ so that they reach $W$ in at most $N$ time units. The flow must occur on the available road network connecting the two facilities. On the one hand, it is desirable that the transport plan utilizes as many different routes as possible so that most of the goods arrive within the prescribed time horizon even in the presence of road congestion, roadwork, etc. On the other hand, it is also important that shorter paths are used to keep the vehicles fuel consumption within a budgetary constraint. 

In this paper, continuing the research initiated in \cite{chen2016networks}, we provide a precise mathematical formulation of the above single commodity network flow problem. Normalizing the mass of goods to one, we formulate a {\em maximum entropy problem} for Markovian distributions on the paths of the network.  The optimal feedback control suitably modifies a prior transition mechanism thereby achieving robustness while limiting the cost. This is accomplished through an appropriate choice of the prior transition involving the {\em adjacency matrix} of the graph. The optimal scheduling, while spreading the mass over all feasible paths, assigns {\em maximum probability} to all {\em minimum cost paths}. 

Differently from the standard literature on controlled Markov chains, the optimal policy ({\em Schr\"{o}dinger bridge}) is {\em not computed through dynamic programming}. The constraint on the final marginal (all the goods should be in the warehouse by day $N$) dictates a different approach. The solution is computed by solving iteratively a Schr\"{o}dinger-Bernstein linear system with nonlinear coupling at the initial and final times. This algorithm, whose convergence was established in \cite{georgiou2015positive}, is related to recent work  in entropy regularization \cite{Cuturi} and equilibrium assignement in economics \cite{GKW} as well as to classical work in statistics \cite{IK}.

Our straightforward  approach also avoids altogether modelling {\em cascading failures} which is a complex and controversial task \cite{savla2014robust}. It is also worthwhile remarking that maximum entropy problems \cite{COVER_THOMAS}
which constitute a powerful inference method, 
find here an alternative use as a tool to produce a desired flow in a network by exploiting the properties of the prior transition mechanism.}

Our intuitive notion of robustness of the routing policy should not be confused with other notions of robustness concerning networks which have been put forward and studied, see e.g. \cite{Albertetal2000,bara2014robustness,olver2010robust,arnoldetal1994,demetrius2005,savla2014robust}.
In particular, in \cite {arnoldetal1994,demetrius2005}, robustness has been defined through a fluctuation-dissipation relation involving the entropy rate. 
This latter notion captures relaxation of a process back to equilibrium after a perturbation and has been used to study both financial and biological networks \cite{Sandhu, Sandhu1}. Our study, inspired by  transportation and data networks, does not concern equilibrium or near equilibrium cases.

{This paper features the following novel contributions:   a) it introduces an explicit index of {\em efficiency} of a transportation plan; b) The choice of the adjacency matrix as prior transition mechanism, which was justified in \cite{chen2016networks} on an intuitive basis, is here further motivated  trough a specific optimization problem; c) we derive an {\em iterated bridge} invariance property; d) we establish the invariance of the {\em most probable paths}. These two invariance properties, which were only briefly mentioned in \cite{chen2016networks} in some special cases, are here fully investigated. Their relevance for transport over networks is also illustrated. e) we study the dependence of the optimal transport on a temperature parameter. The possibility of employing the solution for near-zero temperature as an approximation of the solution to {\em Optimal Mass Transport} (OMT) is also discussed and illustrated through examples.}

The outline of the 
paper is as follows. In Section~\ref{GSB} we introduce 
generalized maximum entropy problems 
In Section \ref{IB} we establish the iterated bridge property, and in Section \ref{MPP} the invariance of the most probable paths. 
Efficiency of a transport policy is introduced in Section~\ref{efficiency}. In Section~\ref{robust}, we introduce robust transport with fixed average path length. Section~\ref{efficient-robust} deals with efficient-robust transportation. In Section~\ref{Tdependence}, the dependence of the optimal transport on the temperature parameter is 
investigated. The results are then illustrated through academic examples in Section~\ref{example}.

 \section{Generalized {maximum entropy problems}}\label{GSB}

We are given a directed, strongly connected (i.e., with a path in each direction between each pair of vertices), aperiodic graph ${\bf G}=(\mathcal X,\mathcal E)$ with vertex set $\mathcal X=\{1,2,\ldots,n\}$ and edge set $\mathcal E\subseteq \mathcal X\times\mathcal X$.  We let  time vary in ${\mathcal T}=\{0,1,\ldots,N\}$, and let ${\mathcal {FP}}_0^N\subseteq\mathcal X^{N+1}$ denote the family of length $N$, feasible paths $x=(x_0,\ldots,x_N)$, namely paths such that $x_ix_{i+1}\in\mathcal E$ for $i=0,1,\ldots,N-1$. 

We seek a probability distribution $\fP$ on ${\mathcal {FP}}_0^N$ with prescribed initial and final marginal probability distributions $\nu_0(\cdot)$ and $\nu_N(\cdot)$, respectively, and such that the resulting random evolution
is closest to a ``prior'' measure $\fM$ on ${\mathcal {FP}}_0^N$ in a suitable sense.
The prior law $\fM$ is induced by the Markovian evolution
 \begin{equation}\label{FP}
\mu_{t+1}(x_{t+1})=\sum_{x_t\in\mathcal X} \mu_t(x_t) m_{x_{t}x_{t+1}}(t)
\end{equation}
with nonnegative distributions $\mu_t(\cdot)$ over $\mathcal X$, $t\in{\mathcal T}$, and weights $m_{ij}(t)\geq 0$ for all indices $i,j\in{\mathcal X}$ and all times. Moreover, to respect the topology of the graph, $m_{ij}(t)=0$ for all $t$ whenever $ij\not\in\mathcal E$.  Often, but not always, the matrix
\begin{equation}\label{eq:matrixM}
M(t)=\left[ m_{ij}(t)\right]_{i,j=1}^n
\end{equation}
does not depend on $t$.
The rows of the transition matrix $M(t)$ do not necessarily sum up to one, so that the ``total transported mass'' is not necessarily preserved.  It occurs, for instance, when $M$ simply encodes the topological structure of the network with $m_{ij}$ being zero or one, depending on whether a certain link exists.
The evolution \eqref{FP} together with the measure $\mu_0(\cdot)$, which we assume positive on $\mathcal X$, i.e.,
\begin{equation}\label{eq:mupositive}
\mu_0(x)>0\mbox{ for all }x\in\mathcal X,
\end{equation}
 induces
a measure $\fM$ on ${\mathcal {FP}}_0^N$ as follows. It assigns to a path  $x=(x_0,x_1,\ldots,x_N)\in{\mathcal {FP}}_0^N$ the value
\begin{equation}\label{prior}\fM(x_0,x_1,\ldots,x_N)=\mu_0(x_0)m_{x_0x_1}\cdots m_{x_{N-1}x_N},
\end{equation}
and gives rise to a flow
of {\em one-time marginals}
\[\mu_t(x_t) = \sum_{x_{\ell\neq t}}\fM(x_0,x_1,\ldots,x_N), \quad t\in\mathcal T.\]
\begin{definition} We denote by ${\mathcal P}(\nu_0,\nu_N)$ the family of probability distributions on ${\mathcal {FP}}_0^N$ having the prescribed marginals $\nu_0(\cdot)$ and $\nu_N(\cdot)$.
\end{definition}

We seek a distribution in this set which is closest to the prior $\fM$ in {\em relative entropy} where, for $P$ and $Q$ measures on $\mathcal X^{N+1}$,  the relative entropy (divergence, Kullback-Leibler index) $\D(P\|Q)$ is
\begin{equation*}
\D(P\|Q):=\left\{\begin{array}{ll} \sum_{x}P(x)\log\frac{P(x)}{Q(x)}, & \support (P)\subseteq \support (Q),\\
+\infty , & \support (P)\not\subseteq \support (Q),\end{array}\right.
\end{equation*}
Here, by definition,  $0\cdot\log 0=0$.
Naturally, while the value of $\D(P\|Q)$ may turn out negative due to miss-match of scaling (in case $Q=\fM$ is not a probability measure), the relative entropy is always jointly convex.
We consider the {\em Schr\"odinger Bridge  Problem} (SBP):

\begin{problem}\label{prob:optimization}
Determine
 \begin{eqnarray}\label{eq:optimization}
\fM^*[\nu_0,\nu_N]:={\rm argmin}\{ \D(P\|\fM) \mid  P\in {\mathcal P}(\nu_0,\nu_N)
\}.
\end{eqnarray}
\end{problem}
\noindent
The following result is a slight generalization (to time inhomogeneous prior) of \cite[Theorem 2.3]{chen2016networks}.
\begin{theorem}\label{solbridge} Assume that the product $M(N-1) M(N-2)\cdots M(1) M(0)$ has all entries positive.
Then there exist nonnegative functions  $\varphi(\cdot)$ and $\hat{\varphi}(\cdot)$ on $[0,N]\times\mathcal X$ satisfying
\begin{subequations}\label{eq:Schroedingersystem}
\begin{eqnarray}\label{Schroedingersystem1}
\varphi(t,i)&=&\sum_{j}m_{ij}(t)\varphi(t+1,j),
\\\hat{\varphi}(t+1,j)&=&\sum_{i}m_{ij}(t)\hat{\varphi}(t,i),\label{Schroedingersystem2}
\end{eqnarray}
for $t\in[0,N-1]$, along with the (nonlinear) boundary conditions
\begin{eqnarray}\label{bndconditions1}
\varphi(0,x_0)\hat{\varphi}(0,x_0)&=&\nu_0(x_0)\\\label{bndconditions2}
\varphi(N,x_N)\hat{\varphi}(N,x_N)&=&\nu_N(x_N),
\end{eqnarray}
\end{subequations}
for $x_0, x_N\in\mathcal X$.
Moreover, the solution $\fM^*[\nu_0,\nu_N]$ to Problem \ref{prob:optimization} is unique and obtained by
\[
\fM^*(x_0,\ldots,x_N)=\nu_0(x_0)\pi_{x_0x_{1}}(0)\cdots \pi_{x_{N-1}x_{N}}(N-1),
\]
where the one-step transition probabilities
\begin{equation}\label{opttransition1}
\pi_{ij}(t):=m_{ij}(t)\frac{\varphi(t+1,j)}{\varphi(t,i)}
\end{equation}
are well defined.
\end{theorem}

The factors $\varphi$ and $\hat{\varphi}$ are unique up to multiplication of $\varphi$ by a positive constant and division of $\hat{\varphi}$ by the same constant.
Let $\varphi(t)$ and $\hat{\varphi}(t)$ denote the column vectors with components $\varphi(t,i)$ and $\hat{\varphi}(t,i)$, respectively, with $i\in\mathcal X$. In matricial form, (\ref{Schroedingersystem1}), (\ref{Schroedingersystem2}
) and (\ref{opttransition1}) read
\begin{equation}
\varphi(t)=M(t)\varphi(t+1),\; ~~\hat{\varphi}(t+1)=M(t)^T\hat{\varphi}(t),
\end{equation}
and
\begin{equation}
	\Pi(t):=[\pi_{ij}(t)]=\diag(\varphi(t))^{-1}M(t)\diag(\varphi(t+1)).
\end{equation}

Historically, the SBP was posed in 1931 by Erwin Schr\"odinger for Brownian particles with a large deviations of the empirical distribution motivation \cite{S1}, see \cite{leo2} for a survey.
The problem was considered in the context of Markov chains and studied in  \cite{pavon2010discrete,georgiou2015positive}, and some generalizations have been discussed in \cite{chen2016networks}. Important connections between SBP and OMT \cite{Vil,AGS,Vil2} have been discovered and developed in \cite{Mik,MT,leo,leo2,CGP1,CGP2,CGP3}.

\subsection{Iterated Bridges}\label{IB}

In this section we explain a rather interesting property of Schr\"odinger bridges which is the following. If, after solving an SBP for a given set of marginals $(\nu_0,\nu_N)$ and a Markovian prior $\fM$ to obtain $\fM^*[\nu_0,\nu_N]$, we decided to update the data $(\nu_0,\nu_N)$ to another set of marginals $(\pi_0,\pi_N)$ then, whether we use as prior $\fM$ or $\fM^*[\nu_0,\nu_N]$ for the SBP with the new marginals $\pi_0$ and $\pi_N$, we obtain precisely the same solution $\fM^*[\pi_0,\pi_N]$. The significance of this property will be discussed later on in the context of robust transportation.


Indeed, take  $\fM^*[\nu_0,\nu_N]$ as prior and consider the corresponding new Schr\"odinger system (in matrix form)
\begin{align*}
&\psi(t)=\Pi(t)\psi(t+1),\; \quad\hat{\psi}(t+1)=\Pi(t)^T\hat{\psi}(t),
\end{align*}
with boundary conditions
\begin{subequations}
\begin{eqnarray}\label{bndconditions1'}
\psi(0,x_0)\hat{\psi}(0,x_0)&=&\pi_0(x_0),\\\label{bndconditions2'}
\psi(N,x_N)\hat{\psi}(N,x_N)&=&\pi_N(x_N).
\end{eqnarray}
\end{subequations}
Note in the above $\Pi(t)=\diag(\varphi(t))^{-1}M(t)\diag(\varphi(t+1))$, therefore, it can be written as
\begin{subequations}\label{NSS}
\begin{align}
&\hspace*{-7pt}\diag(\varphi(t))\psi(t)=M(t)\diag(\varphi(t+1))\psi(t+1),\\
&\hspace*{-7pt}\diag(\varphi(t+1))^{-1}\hat{\psi}(t+1)=M(t)^T\diag{\varphi(t))}^{-1}\hat{\psi}(t).
\end{align}
\end{subequations}
The new transition matrix $Q^*$ is given by
\begin{align*}
	Q^*(t)&=\diag(\psi(t))^{-1}\Pi(t)\diag(\psi(t+1))\\&=\diag(\psi(t))^{-1}\diag(\varphi(t))^{-1}\\
	&\hspace*{5pt}\times M(t)\diag(\varphi(t+1))\diag(\psi(t+1)).
\end{align*}
Let $\psi_1(t)=\diag(\varphi(t))\psi(t)$ and $\hat{\psi}_1(t)=\diag(\varphi(t))^{-1}\hat{\psi}(t)$, then
\[Q^*(t)=\diag(\psi_1(t))^{-1}M(t)\diag(\psi_1(t+1)).
\]
By (\ref{NSS}), $\psi$ and $\hat{\psi}$ are vectors with positive components satisfying
\[
\psi_1(t)=M(t)\psi_1(t+1),\; ~~\hat{\psi}_1(t+1)=M(t)^T\hat{\psi}_1(t).
\]
Moreover, they satisfy the boundary conditions
\begin{subequations}
\begin{eqnarray}\label{bndconditions1''}
\psi_1(0,x_0)\hat{\psi}_1(0,x_0)&=&\pi_0(x_0)\\\label{bndconditions2''}
\psi_1(N,x_N)\hat{\psi}_1(N,x_N)&=&\pi_N(x_N).
\end{eqnarray}
\end{subequations}
Thus, $(\psi_1,\hat{\psi}_1)$ provide the solution to Problem \ref{prob:optimization} when $\fM$ is taken as prior.

Alternatively, observe the transition matrix $Q^*(t)$ resulting from the two problems is the same and so is the initial marginal. Hence,   the solutions of the SBP with marginals $\pi_0$ and $\pi_N$ and prior transitions $\Pi(t)$ and $M(t)$ are identical.

Thus, ``the bridge over a bridge over a prior" is the same as the ``bridge over the prior,'' i.e., iterated bridges produce the same result.
It is should be observed that this result for probability distributions is not surprising since the solution is in the same {\em reciprocal class} as the prior (namely, it has the same three times transition probability), cf.\ \cite{Jam74,levy1990modeling,WC}. It could then be described as the fact that only the reciprocal class of the prior matters; this is can be seen from Schr\"{o}dinger's original construction \cite{S1}, and also \cite[Section III-B]{georgiou2015positive} for the case of Markov chains. This result, however, is more general since the prior is not necessarily a probability measure. 
 
 In information theoretic terms, the bridge (i.e., probability law on path spaces) corresponding to $Q^*$ is the I-projection
in the sense of Cziszar \cite{csiszar} of the prior onto the set of measures that are consistent with the initial-final marginals. The above result, however, is not simply an ``iterated information-projection" property, since  $\fM^*[\nu_0,\nu_N]$ is the I-projection of $\fM$ onto $\mathcal P(\nu_0,\nu_N)$ which does not contain $\mathcal P(\pi_0,\pi_N)$ being in fact disjoint from it.

\subsection{Invariance of most probable paths}\label{MPP}

{Building on the {\em logarithmic transformation} of Fleming, Holland, Mitter and others, the connection between SBP and stochastic control was developed from the early nineties on \cite{Dai91,blaquiere1992controllability,DaiPav90,PavWak91}.  More recently Brockett studied steering of the Louiville equation \cite{brockett2012notes}.} In  \cite[Section 5]{Dai91}, Dai Pra established an interesting path-space  property of the Schr\"{o}dinger bridge for diffusion processes, that the ``most probable path" \cite{DB,TW} of the prior and the solution are the same. Loosely speaking, a most probable path is similar to a {\em mode} for the path space measure $P$.  More precisely, if both drift $b(\cdot,\cdot)$ and diffusion coefficient $\sigma(\cdot,\cdot)$ of the Markov diffusion process
\[dX_t=b(X_t,t)dt+ \sigma(X_t,t)dW_t
\]
are smooth and bounded, with $\sigma(x,t)\sigma(x,t)^T>\eta I$, $\eta>0$, and $\{x(t) \mid 0\le t\le T\}$ is a path of class $C^2$, then there exists an asymptotic estimate of the probability $P$ of a small tube around $x(t)$ of radius $\epsilon$. It  follows from this estimate that the most probable path is the minimizer in a deterministic calculus of variations problem where the Lagrangian is an {\em Onsager-Machlup functional}, see \cite[p.\ 532]{IW} for the full story{\footnote {The Onsager-Machlup functional was introduced in {\cite{OM} to develop a theory of fluctuations in equilibrium and nonequilibrium thermodynamics.} }.

The concept of most probable path is, of course, much less delicate in our discrete setting. We define it for general positive measures on paths. Given a positive measure $\fM$ as in Section \ref{GSB} on the feasible paths of our graph $\bf G$, we say that $x=(x_0,\ldots,x_N)\in {\mathcal {FP}}_0^N$ is of {\em maximal mass} if for all other feasible paths $y\in{\mathcal {FP}}_0^N$ we have $\fM(y)\le\fM(x)$. Likewise we consider paths of {\em maximal mass} connecting particular nodes. It is apparent that paths of maximal mass always exist but are, in general, not unique. If $\fM$ is a probability measure, then the maximal mass paths - most probable paths are simply the modes of the distribution. We establish below that the maximal mass paths joining two given nodes under the solution of a Schr\"odinger Bridge problem as in Section \ref{GSB} are the same as for the prior measure.

\begin{prop} \label{invariance}
Consider marginals $\nu_0$ and $\nu_1$ in Problem \ref{prob:optimization}. Assume that $\nu_0(x)>0$ on all nodes 
$x\in\mathcal X$ and that the product $M(N-1)\cdot M(N-2)\cdots M(1)\cdot M(0)$ of transition probability matrices of the prior has all positive elements (cf.\ with $M$'s as in \eqref{eq:matrixM}). Let  $x_0$ and $x_N$ be any two nodes. Then, under the solution $\fM^*[\nu_0,\nu_N]$ of the SBP, the family of maximal mass paths joining $x_0$ and $x_N$ in $N$ steps is the same as under the prior measure $\fM$.

\noindent
\proof
{Suppose path $y=(y_0=x_0,y_1,\ldots, y_{N-1},y_N=x_N)$ has maximal mass under the prior $\fM$. In view of (\ref{prior}) and (\ref{opttransition1}) and assumption (\ref{eq:mupositive}), we have
\begin{eqnarray}\nonumber\fM^*[\nu_0,\nu_N](y)&=&\nu_0(y_0)\pi_{y_0y_1}(0)\cdots \pi_{y_{N-1}y_N}(N-1)\\
&=&\frac{\nu_0(x_0)}{\mu_0(x_0)}\frac{\varphi(N,x_N)}{\varphi(0,x_0)}\fM(y_0,y_1,\ldots,y_N).\nonumber
\end{eqnarray}
Since the quantity
\[\frac{\nu_0(x_0)}{\mu_0(x_0)}\frac{\varphi(N,x_N)}{\varphi(0,x_0)}
\]
is positive and does not depend on the particular path joining $x_0$ and $x_N$, the conclusion follows.}
\qed

\end{prop}
The calculation in the above proof actually establishes the following stronger result. 
\begin{prop} Let  $x_0$ and $x_N$ be any two nodes in $\mathcal X$.
Then, under the assumptions of Proposition \ref{invariance}, the measures $\fM$ and $\fM^*[\nu_0,\nu_N]$, restricted on the set of paths that begin at $x_0$ at time $0$ and end at $x_N$ at time $N$, are identical.
\end{prop}

\section{Robust transport}

In this section, we first discuss notions of efficiency of a transportation plan and then introduce entropy as a surrogate for robustness.

\subsection{Efficiency of a transport plan}\label{efficiency}
Inspired by the celebrated paper  \cite{watts1998collective}, we introduce below a measure of {\em efficiency} of a  {\em transportation plan over a certain finite-time horizon and a given network}.

For the case of undirected and connected graphs,
 {\em small-world networks} \cite{watts1998collective} were identified as networks being highly clustered but with small characteristic path length $L$, where 
\[L:= 
\frac{1}{n(n-1)}\sum_{i\neq j}d_{ij}
\]
and $d_{ij}$ is the shortest path length between vertices $i$ and $j$. The inverse of the characteristic path length $L^{-1}$ is an index of efficiency  of ${\bf G}$. There are other such indexes, most noticeably the global efficiency $E_{{\rm glob}}$ introduced in \cite{latora2001efficient}. This is defined as $E_{{\rm glob}}=E({\bf G})/E({\bf G}_{\rm id})$ where
\[ E({\bf G})=\frac{1}{n(n-1)}\sum_{i\neq j}\frac{1}{d_{ij}}
\]
and ${\bf G}_{\rm id}$ is the complete network with all possible edges in place.  Thus, $0\le E_{{\rm glob}}\le 1$.
However, as argued on \cite[p. 198701-2]{latora2001efficient}, it is $1/L$ which ``measures the efficiency of a sequential system (i.e., only one packet of information goes along the network)". $E_{{\rm glob}}$, instead, measures the efficiency of a parallel system, namely one in which all nodes  concurrently exchange packets of information.  Since we are interested in the efficiency of a specific transportation plan, we define below efficiency by a suitable adaptation of the index $L$.

Consider a strongly connected, aperiodic, directed graph ${\bf G}=(\mathcal X,\mathcal E)$ as in Section \ref{GSB}.  To each edge $ij$ is now associated a length $l_{ij}\ge 0$. If $ij\not\in\mathcal E$, we set $l_{ij}=+\infty$. The length may represent distance, cost of transport/communication/etc. Let ${\mathcal T}=\{0,1,\ldots,N\}$ be the time-indexing set. For a path $x=(x_0,\ldots,x_N)\in\mathcal X^{N+1}$, we define the length of $x$ to be
\[l(x)=\sum_{t=0}^{N-1}l_{x_tx_{t+1}}.
\]
We consider the situation where initially at time $t=0$ the mass is distributed on $\mathcal X$ according to $\nu_0(x)$ and needs to be distributed according to $\nu_N(x)$ at the final time $t=N$.           These masses are normalized to sum to one, so that they are probability distributions.  A transportation plan $P$ is a probability measure on the (feasible) paths of the network having the prescribed marginals $\nu_0$ and $\nu_N$ at the initial and final time, respectively. A natural adaptation of the characteristic path length is to consider the {\em average path length of the transportation plan} $P$, which we define as
\begin{equation}\label{internal}
L(P)=\sum_{x\in\mathcal X^{N+1}}l(x)P(x)
\end{equation}
with the usual convention $+\infty\times 0=0$. This is entirely analogous to a thermodynamic quantity, the {\em internal energy}, which is defined as the expected value of the Hamiltonian observable in state $P$. Clearly, $L(P)$ is finite if and only if the transport takes place on actual, existing links of ${\bf G}$. Moreover, only the paths which are in the {\em support} of $P$ enter in the computation of $L(P)$. One of the goals of a transportation plan is of course to have small average path length since, for instance, cost might simply be proportional to length. Determining the probability measure that minimizes \eqref{internal} can be seen to be an OMT problem.

\subsection
{Problem formulation}\label{robust}

 Besides efficiency, another desirable property of a transport strategy is to ensure robustness with respect to links/nodes failures, the latter being due possibly to malicious attacks. We therefore seek a transport plan in which the mass spreads, as much as it is allowed by the network topology, before reconvening at time $t=N$ in the sink nodes. We achieve this by selecting a transportation plan $P$ that has a suitably high entropy $S(P)$, where
\begin{equation}\label{eq:entropy}
S(P)=-\sum_{x\in \mathcal X^{N+1}}P(x)\ln P(x).
\end{equation}
Thus, in order to attain a level of robustness while guaranteeing a relatively low average path length (cost), we formulate below a constrained optimization problem that weighs in both $S(P)$ as well as $L(P)$.


We begin by letting $\bar{L}$ designate a suitable bound on the average path length (cost) that we are willing accept. Clearly, we need that
\begin{subequations}\label{lengthbounds}
\begin{equation}
l_m:=\min_{x\in\mathcal X^{N+1}} l(x)\le \bar{L}.
\end{equation}
We will also assume that
\begin{equation}
\bar{L}\le \frac{1}{|{\mathcal {FP}}_0^N|}\sum _{x\in{\mathcal {FP}}_0^N}l(x).
\end{equation}
\end{subequations}
The rationale behind the latter, i.e., requiring an upper bound as stated, will be explained in Proposition \ref{properties} below.

Let $\mathcal P$ denote the family of probability measures on $\mathcal X^{N+1}$. The probability measure that maximizes the entropy $S(P)$ subject to a path-length constraint $L(P)=\bar L$ is the Boltzmann distribution
\begin{equation}\label{Boltz}
\hspace*{-7pt}P_T^*(x)=Z(T)^{-1}\exp[-\frac{l(x)}{T}],\; Z(T)=\sum_x\exp[-\frac{l(x)}{T}],
\end{equation}
where the parameter (temperature) $T$ depends on $\bar L$. To see this, 
%
consider the Lagrangian
\begin{equation}\label{lagrangian}{\cal L}(P,\lambda):=S(P)+\lambda(\bar{L}- L(P)),
\end{equation}
and observe that the Boltzman distribution \eqref{Boltz} satisfies the first order optimality condition of $\cL$ with $T=1/\lambda$.
Clearly, the Boltzmann distribution has support on the feasible paths ${\mathcal {FP}}_0^N$.
Hence, we get a 
version of Gibbs' variational principle that the Boltzmann distribution $P_T^*$ minimizes the {\em free energy} functional
\begin{equation}
\label{eq:F}
F(P,T):=L(P)-TS(P)
\end{equation}
over $\mathcal P$. An alternative way to establish the minimizing property of the Boltzmann's distribution is to observe that
\begin{equation}
\label{eq:F2}
F(P,T)=T\D(P\|P_T^*)-T\log Z,
\end{equation}
and therefore, minimizing the free energy over $\mathcal P$ is equivalent to minimizing the relative entropy $\D(P\|P_T^*)$ over $ P\in\mathcal P$,
which ensures that the minimum is unique. The following properties of $P_T^*$ are noted, see e.g. \cite[Chapter 2]{MicheleNotes}.

\begin{prop} \label{properties} The following hold:
\begin{itemize}
\item[i)] For $T\nearrow+\infty$, $P_T^*$ tends to
the uniform distribution on all feasible paths.
\item[ii)] For $T\searrow 0$, $P_T^*$ tends to concentrate on the set of feasible paths having minimal length.
\item[iii)] Assuming that $l(\cdot)$ is not constant over ${\mathcal {FP}}_0^N$ then, for each value $\bar{L}$ satisfying the bounds (\ref{lengthbounds}), there exists a unique nonnegative value of $T=\lambda^{-1}\in [0, \infty]$ such that $P_T^*$ maximizes $S(P)$ subject to $L(P)=\bar L$.
\end{itemize}
\end{prop}

We also
observe the Markovian nature of the measure $P_T^*$. 
Indeed, recall that a positive measure $\fM$ on $\mathcal X^{N+1}$ is Markovian if it can be expressed as in (\ref{prior}).
Since
\begin{equation}\label{Bolt}P_T^*(x_0,x_1,\ldots,x_N)=Z(T)^{-1}\prod_{t=0}^{N-1}\exp[-\frac{l_{x_tx_{t+1}}}{T}],
\end{equation}
which is exactly in the form (\ref{prior}), we conclude that $P_T^*$ is ({\em time-homogeneous}) Markovian  with uniform initial measure $\mu(x_0)\equiv Z(T)^{-1}$  and time-invariant transition matrix given by
\begin{equation}\label{transition}                                                                 M_T=\left[\exp\left(-\frac{l_{ij}}{T}\right)\right]_{i,j=1}^n.
\end{equation}
Observe however that, in general, $M_T$ is not stochastic (rows do not sum to one). Moreover, observe that, after suitable normalization, $M_T$ represents the transition matrix of a chain where probabilities of transition between nodes are inversely proportional to the length of the links.

Consider now $\nu_0$ and $\nu_N$ distributions on $\mathcal X$.  These are the ``starting'' and ``ending'' concentrations of resources for which we seek a transportation plan. We denote by ${\mathcal P}(\nu_0,\nu_N)$ the family of probability distributions on paths $x\in\mathcal X^{N+1}$ having $\nu_0$ and $\nu_N$ as initial and final marginals, respectively, and we
consider the problem to maximize the entropy subject to marginal and length constraints:
\begin{problem}\label{prob3} Maximize $S(P)$ subject to $P\in\mathcal P(\nu_0,\nu_N)$ and
$L(P)=\bar{L}$.   
\end{problem}

Note that the solution to Problem \ref{prob3} depends on $\bar L$ as well as the two marginals $\nu_0, \nu_N$ and that when $\bar L$ is too close to $l_m$, the problem may be infeasible.

Once again, bringing in the Lagrangian (\ref{lagrangian}), which now needs to be minimized over $\mathcal P(\nu_0,\nu_N)$, we see that Problem \ref{prob3} is equivalent to solving the following {\em Schr\"{o}dinger Bridge} problem for a suitable value of the parameter $T$.}

\begin{problem}\label{prob4}
{\rm minimize} \quad $\{\D(P\|P_T^*)\mid P\in\mathcal P(\nu_0,\nu_N)\}$.
\end{problem}

Thus, {\em employing path space entropy as a measure of robustness},  the solution to Problem \ref{prob4}, denoted by $\fM^*_T(\nu_0,\nu_N)$  and constructed in accordance with Theorem \ref{solbridge}, minimizes a suitable {\em free energy functional} with the temperature parameter 
specifying the tradeoff between
efficiency and robustness.
 Thus, Problem \ref{prob4} can be viewed as an SBP as in Section \ref{GSB} where the ``prior" measure $P_T^*$ is Markovian.

\section
{Structure of robust transport}\label{efficient-robust}

We now address in detail
Problem \ref{prob4}, namely, to identify a probability distribution $P$ on ${\mathcal {FP}}_0^N$ that minimizes $\D(\cdot\|P_T^*)$ over $\mathcal P(\nu_0,\nu_N)$ where $P^*_T$ is the Boltzmann distribution (\ref{Bolt})--the minimizing law being denoted by $\fM^*_T[\nu_0,\nu_N]$ as before.
We show below that the two invariant properties discussed in the previous two sessions can be used to determine 
an optimal
 transport policy. We also show that the $\fM^*_T[\nu_0,\nu_N]$ inherits from the Boltzmann distribution $P^*_T$ properties as dictated by Proposition \ref{properties}.

Initially, for simplicity, we consider the situation where at time $t=0$ the whole mass is concentrated on node $1$ ({\em source}) and at time $t=N$ it is concentrated on node $n$ ({\em sink}), i.e., $\nu_0(x)=\delta_1(x)$ and $\nu_N(x)=\delta_n(x)$. We want to allow (part of) the mass to reach the end-point ``sink'' node, if this is possible, in less than $N$ steps and then remain there until $t=N$. In order to ensure that is possible, we assume that there exists a self-loop at node $n$, i.e., $M_{Tnn}>0$.  Clearly, $\fM^*_T(\delta_1,\delta_n)(\cdot)=P^*_T[\cdot| Y_0=1, Y_N=n]$. The Schr\"{o}dinger bridge theory provides transition probabilities so that,
for a path $y=(y_0,y_1,\ldots,y_N)$,

{\footnotesize
\begin{align}
\fM^*_T(\delta_1,\delta_n)(y)&=\delta_1(y_0)\prod_{t=1}^{N-1} \exp\left(-\frac{l_{y_ty_{t+1}}}{T}\right)\!\!\nonumber\frac{\varphi_T(t+1,y_{t+1})}{\varphi_T(t,y_t)}\\
&=\delta_1(y_0)\frac{\varphi_T(N,y_N)}{\varphi_T(0,y_0)}\left[\exp\left(-\frac{l(y)}{T}\right)\right],\label{optrobeff}
\end{align}
}
cf.\  (\ref{prior}) and (\ref{opttransition1}).
Here $l(y)=\sum_{t=0}^{N-1}l_{y_ty_{t+1}}$ is the length of path $y$ and $\varphi_T$ satisfies together with $\hat{\varphi}_T$ the Schr\"{o}dinger system (\ref{eq:Schroedingersystem}) with $m_{ij}(t)=\exp\left(-\frac{l_{ij}}{T}\right)$ and $\nu_0(x)=\delta_1(x), \nu_N(x)=\delta_n(x)$.

In \cite[Section VI]{chen2016networks}, Problem \ref{prob4} was first studied with a prior measure $\fM_l$ having certain special properties. To introduce  this particular measure, we first recall (part of) a fundamental result from linear algebra \cite{horn2012matrix}.

\begin{theorem}[Perron-Frobenius]\label{perrontheorem}
Let $A=\left(a_{ij}\right)$ be an $n\times n$ matrix with nonnegative entries. Suppose there exists $N$ such that $A^N$ has only positive entries,
and let $\lambda_A$ be its spectral radius. Then\\[-.25in]
\begin{enumerate}
\item[i)] $\lambda_A>0$ is an eigenvalue of $A$;
\item[ii)] $\lambda_A$ is a simple eigenvalue;
\item[iii)] there exists an eigenvector $v$ corresponding to $\lambda_A$ with strictly positive entries.
\end{enumerate}
\end{theorem}

Consider now the weighted adjacency matrix $B=M_T$ in (\ref{transition}) 
(where we dropped the subscript $T$ as it will be fixed throughout this section). Assume that $B^N$ has all positive elements so that we can apply the Perron-Frobenius theorem. Let $u$ and $v$ be the left and right eigenvectors with positive components of the matrix $B$ corresponding to the spectral radius $\lambda_B$. We have
\begin{equation}\label{eigen}B^Tu=\lambda_Bu, \quad Bv=\lambda_Bv.
\end{equation}
We assume throughout that $u$ and $v$ are chosen so that $\sum_iu_iv_i=1$.
Then, for $y_0=i$ and $y_t=j$, define
\begin{equation}
\fM_l(i,y_1,\ldots,\y_{t-1},j):=\lambda_B^{-t}u_iv_je^{-\sum_{k=0}^{t-1}l_{y_ky_{k+1}}}.
\end{equation}
The corresponding transition matrix is
\begin{equation}R_l=\lambda_B^{-1}\diag(v)^{-1}B\diag(v).
\end{equation}
It admits the {\em invariant measure}
\begin{equation}\label{IM}
\mu_l(i)=u_i v_i.
\end{equation}
Note that $\fM_l$ and the Boltzmann distribution $P_T^*$ have the same transition matrix but different initial distributions. In \cite{chen2016networks}, to which we refer for motivation and more details, the following problem was studied. \begin{problem}\label{prob5}
{\rm minimize} \quad $\{\D(P\|\fM_l)\mid P \in \mathcal P(\nu_0,\nu_N)\}$.
\end{problem}

Under the assumption that $B^N$ has all positive entries, this Schr\"{o}dinger bridge problem has a unique solution $\fM^*_l$. In \cite[Theorem 3.4]{chen2016networks}, it was also shown that $\fM_l$ is itself the solution of a Schr\"{o}dinger bridge problem with equal marginals and the Boltzmann distribution (\ref{Boltz}) as prior. Thus, by the iterated bridge property of Section \ref{IB}, $\fM^*_l$ coincides with the solution of Problem \ref{prob4} for any choice of the initial-final marginals $\nu_0$ and $\nu_N$.

We recall the following rather surprising result \cite[Theorem 6.1]{chen2016networks} which includes the invariance of the most probable paths in Problem \ref{prob4} (Proposition \ref{invariance}).
\begin{theorem} $\fM^*_l$ gives equal probability to paths $y\in\mathcal X^{N+1}$ of equal length between any two given nodes. In particular, it assigns maximum and
equal probability to minimum length paths. 
\end{theorem}

This result is relevant when the solution of Problem \ref{prob4} for low temperature is used as an approximation to OMT, see Remark \ref{remark} in the next section.
Finally, an important special case occurs when $l_{ij}=0$ for existing links and $+\infty$ for non-existing. Then the matrix $B$ reduces to the unweighted adjacency matrix $A$ and the measure $\fM_l$ to the so-called Ruelle-Bowen random walk $\fM_{RB}$. The only concern in the transport policy is in maximizing path family entropy to achieve robustness, see \cite[Sections 4 and 5]{chen2016networks} for details.

\section{Dependence of
robust transport on $T$}\label{Tdependence}
Below we study how the solution $\fM^*_T[\delta_{x_0},\delta_{x_N}]$ to  Problem \ref{prob4}  varies with the temperature parameter $T$. Here, $x_0$, $x_N$ are specified nodes where mass is concentrated at the start and end times, and $\delta_{x^\prime}(x)=1$ when $x=x^\prime$ and zero otherwise. It should be noted that similar results hold for general marginal distributions as well, which are not necessarily Dirac.

\begin{theorem} \label{temperature}Consider the solution $\fM^*_T[\delta_{x_0},\delta_{x_N}]=:\fM^*_T$ to {\em Problem \ref{prob4}} with $\nu_0(x)=\delta_{x_0}(x)$ and $\nu_N(x)=\delta_{x_N}(x)$. Let $l_m(x_0,x_N)=\min_{y\in\mathcal X^{N+1}(x_0,x_N)} l(y)$, i.e., the minimum length of $N$-step paths originating in $x_0$ and terminating in $x_N$. Then

\begin{itemize}
\item[i)] For $T\searrow 0$, $\fM^*_T$ tends to concentrate itself on the set of feasible, minimum length paths joining $x_0$ and $x_N$ in $N$ steps. Namely, if $y=(y_0=x_0,y_1,\ldots,y_{N-1},y_N=x_N)$ is such that $l(y)>l_m(x_0,x_N)$, then $\fM^*_T(y)\searrow 0$ as $T\searrow 0$.
\item[ii)] For $T\nearrow+\infty$, $\fM^*_T$ tends to the uniform distribution on all feasible paths joining $x_0$ and $x_N$ in $N$ steps.
\item[iii)] Suppose $\mathcal X^{N+1}(x_0,x_N)$ is not a singleton and that $l(\cdot)$ is not constant over it. Then, for each value $\bar{L}$ satisfying the bounds
\begin{equation}\label{lengthboundsSB} \nonumber
l_m(x_0,x_N)\le \bar{L}\le \frac{1}{|\mathcal X^{N+1}(x_0,x_N)|}\sum _{y\in\mathcal X^{N+1}(x_0,x_N)}l(y)
\end{equation}
there exists a unique value of $T\in [0,+\infty]$ such that $\fM^*_T$ satisfies the constraint 
$L(\fM^*_T)=\bar{L}$ and therefore solves  Problem \ref{prob3}.
\end{itemize}
\end{theorem}

\proof
Observe first that, since $\fM^*_T$ is a probability measure on $\mathcal X^{N+1}$, it must satisfy by (\ref{optrobeff})
{\footnotesize
\begin{align}\nonumber
1=\sum_{y\in \mathcal X^{N+1}}\fM^*_T(y)&=\sum_{y\in \mathcal X^{N+1}}\delta_1(y_0)\frac{\varphi_T(N,y_N)}{\varphi_T(0,y_0)}\left[\exp\left(-\frac{l(y)}{T}\right)\right]\\&\hspace*{-1.6cm}=
\sum_{y\in \mathcal X^{N+1}(x_0,x_N)}\delta_1(y_0)\frac{\varphi_T(N,x_N)}{\varphi_T(0,x_0)}\left[\exp\left(-\frac{l(y)}{T}\right)\right],\label{normalization}
\end{align}
}
where we have used the fact that the initial and final marginals of $\fM^*_T$ are $\delta_1$ and $\delta_n$, respectively. It follows that
{\footnotesize
\begin{align} \nonumber
\frac{\varphi_T(0,x_0)}{\varphi_T(N,x_N)}&=\sum_{y\in \mathcal X^{N+1}(x_0,x_N)}\delta_1(y_0)\left[\exp\left(-\frac{l(y)}{T}\right)\right]\\
&=\sum_{y\in \mathcal X^{N+1}(x_0,x_N)}\left[\exp\left(-\frac{l(y)}{T}\right)\right],\label{generalizedpartition}
\end{align}
}
where again $\mathcal X^{N+1}(x_0,x_N)$ denotes the family of paths joining $x_0$ and $x_N$ in $N$ time periods.

\noindent
{\em Proof of i):} Let $y=(y_0=x_0,y_1,\ldots,y_{N-1},y_N=x_N)$ be such that $l(y)>l_m(x_0,x_N)$. Then
\[\fM^*_T(y)=\frac{\varphi_T(N,x_N)}{\varphi_T(0,x_0)}\left[\exp\left(-\frac{l(y)}{T}\right)\right].
\]
By (\ref{generalizedpartition}), we have
$\frac{\varphi_T(0,x_0)}{\varphi_T(N,x_N)}\ge \exp\left(-\frac{l_m(x_0,x_N)}{T}\right)$.
Hence,
\begin{align*}
\fM^*_T(y)&=\frac{\varphi_T(N,x_N)}{\varphi_T(0,x_0)}e^{-\frac{l(y)}{T}} \le e^{-\frac{l(y)-l_m(x_0,x_N)}{T}}.
\end{align*}
Since $l(y)-l_m(x_0,x_N)>0$, the right-hand side  tends to zero as $T\searrow 0$.

\noindent
{\em Proof of ii):} For $T\nearrow +\infty$, $\exp\left(-\frac{l(y)}{T}\right)$ tends to $1$ for all paths $y\in\mathcal X^{N+1}(x_0,x_N)$. Since $\frac{\varphi_T(N,x_N)}{\varphi_T(0,x_0)}$ does not depend on the specific path in $\mathcal X^{N+1}(x_0,x_N)$ (it is just a normalization like the partition function), we conclude that  as $T$ tends to infinity, $\fM^*_T$ tends to the uniform distribution on $\mathcal X^{N+1}(x_0,x_N)$.

\noindent
{\em Proof of iii):} Note that Problem \ref{prob3} is feasible when $l_m(x_0,x_N)\le \bar{L}$ 
holds. By standard Lagrangian duality theory, there exists a Lagrangian multiplier $\lambda\in [0, \infty]$ such that the maximizer of the corresponding Lagrangian \eqref{lagrangian} over $\mathcal P(\nu_0,\nu_N)$ is the solution of Problem \ref{prob3}\footnote{Actually, using (\ref{generalizedpartition}), it is easy to see that $L(\fM^*_T)= \E_{\fM^*_T}[l(Y)]$ is a strictly increasing function of $T$. Indeed, \[\frac{\partial \E_{\fM^*_T}[l(Y)]}{\partial T}=\frac{1}{T^2}\var_{\fM^*_T}[l(Y)]\] where $l(Y)=\sum_{t=0}^{N-1}l_{Y_tY_{t+1}}$ and $Y=(Y_0, Y_1,\ldots,Y_N)$ is the Markov chain. In view of Points $1$ and $2$, we conclude that  $\E_{\fM^*_T}[l(Y)]$ bijectively maps $[0,+\infty]$ onto
\[
[l_m(x_0,x_N),\frac{1}{|\mathcal X^{N+1}(x_0,x_N)|}\sum _{y\in\mathcal X^{N+1}(x_0,x_N)}l(y)].
\]}.
On the other hand, maximizing \eqref{lagrangian} over $\mathcal P(\nu_0,\nu_N)$ is equivalent to solving Problem \ref{prob4} with $T=1/\lambda$. This completes the proof.
\qed

\begin{remark} \label{remark}
{Let us interpret $l_{ij}$ as the cost of transporting a unit mass over the link $ij$.  Then $L(P)$ is the {\em expected cost} corresponding to the transport plan $P$. For $T=0$, the free energy functional reduces to $L(P)$ as our problem amounts to a discrete OMT problem \cite{rachev1998mass}. In this, one seeks minimum cost paths --a combinatorial problem which can also be formulated as a linear programming problem \cite{bazaraa2011linear}.
Precisely as in the diffusion case \cite{CGP1,CGP2,CGP3}, we also see that when the ``heat bath" temperature is close to $0$, the solution of the Schr\"{o}dinger bridge problem is close to the solution of the discrete OMT problem (claim i) of Theorem \ref{temperature}). Since for the former an efficient iterative algorithm is available \cite{georgiou2015positive}, we see that also in this discrete setting the SBP provides a valuable computational approach to solving OMT problems. This is illustrated in the next section through an academic example. It should also be observed that the measure $\fM^*_T(\delta_1,\delta_n)$ is just a ``Boltzmann measure" on the subset of $\mathcal X^{N+1}$ of paths originating in $x_0$ and terminating in $x_N$. 
Thus the above proof is analogous to the classical one for $P^*_T$.}
\end{remark}
\section{Examples}\label{example}

\begin{figure}[h]
\begin{center}
\includegraphics[width=0.42\textwidth]{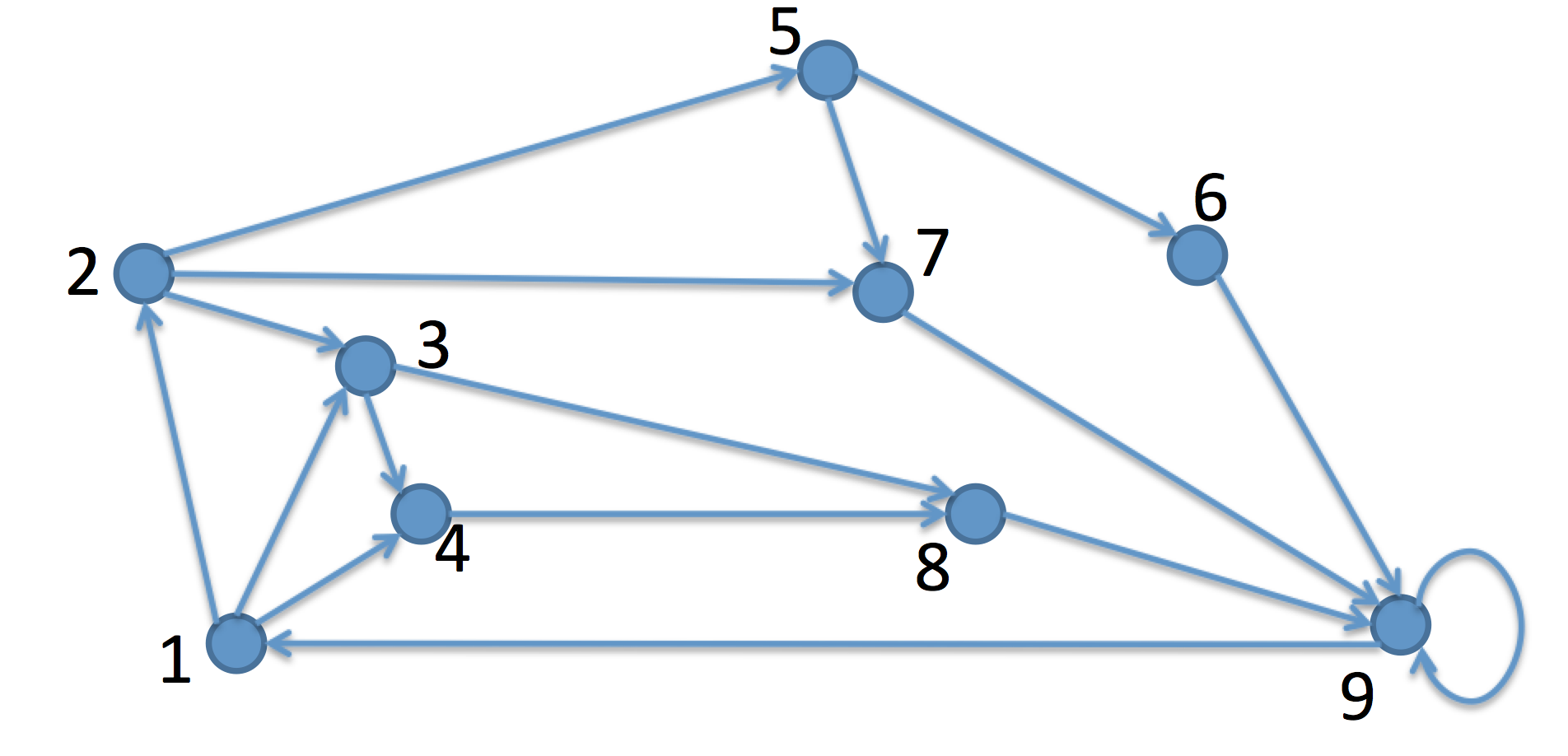}
\caption{Network topology}
\label{fig:graph}
\end{center}
\end{figure}

Consider the graph in Figure \ref{fig:graph}. 
We seek to transport a unit mass from node $1$ to node $9$ in $N =3$ and $4$ steps. We first consider the case where the costs of all the edges are equal to $1$. Here we add a zero cost self-loop at $9$, i.e., $l_{99}=0$. The shortest path from node $1$ to $9$ is of length $3$ and there are three such paths, which are
$1-2-7-9$, $1-3-8-9$ and $1-4-8-9$. If we want to transport the mass with a minimum number of  steps, we may end up using one of these three paths. To achieve robustness, we apply the Schr\"odinger bridge framework. Since all the three feasible paths have equal length, we get a transport plan with equal probabilities using all these three paths, regardless of the choice of temperature $T$. The evolution of mass distribution is given by
 \[
      \tiny
       \left[
       \begin{matrix}
       1 & 0 & 0 & 0 & 0 & 0 & 0 & 0 & 0\\
       0 & 1/3 & 1/3 & 1/3 & 0 & 0 & 0 & 0 & 0\\
       0 & 0 & 0 & 0 & 0 & 0 & 1/3 & 2/3 & 0\\
       0 & 0 & 0 & 0 & 0 & 0 & 0 & 0 & 1
       \end{matrix}
       \right],
\]
where the four rows of the matrix show the mass distribution at time step $t=0, 1, 2 ,3$ respectively. As we can see, the mass spreads out first and then goes to node $9$. When we allow for more steps $N=4$, the mass spreads even more before reassembling at node $9$, as shown below, for $T=1$,
 \[
       \tiny
       \left[
       \begin{matrix}
       1 & 0 & 0 & 0 & 0 & 0 & 0 & 0 & 0\\
       0 & 0.4705 & 0.3059 & 0.2236 & 0 & 0 & 0 & 0 & 0\\
       0 & 0 & 0.0823 & 0.0823 & 0.1645 & 0 & 0.2236 & 0.4473 & 0\\
       0 & 0 & 0 & 0 & 0 & 0.0823 & 0.0823 & 0.1645 & 0.6709\\
       0 & 0 & 0 & 0 & 0 & 0 & 0 & 0 & 1
       \end{matrix}
       \right].
\]
There are $7$ feasible paths of length $4$, which are $1-2-7-9-9$, $1-3-8-9-9$, $1-4-8-9-9$, $1-2-5-6-9$, $1-2-5-7-9$, $1-3-4-8-9$ and $1-2-3-8-9$. The amount of mass traveling along these paths are 
	\[
		0.2236, 0.2236, 0.2236, 0.0823, 0.0823, 0.0823, 0.0823.
	\]
The first three are the most probable paths. This is consistent with Proposition \ref{invariance} since they are the paths with minimum length. If we change the temperature $T$, the flow changes. The set of most probable paths, however, remains invariant. In particular, when $T=0.1$, the flow concentrates on the most probable set (effecting OMT-like transport), as shown below
 \[
       \tiny
       \left[
       \begin{matrix}
       1 & 0 & 0 & 0 & 0 & 0 & 0 & 0 & 0\\
       0 & 0.3334 & 0.3333 & 0.3333 & 0 & 0 & 0 & 0 & 0\\
       0 & 0 & 0 & 0 & 0 & 0 & 0.3334 & 0.6666 & 0\\
       0 & 0 & 0 & 0 & 0 & 0 & 0 & 0 & 1\\
       0 & 0 & 0 & 0 & 0 & 0 & 0 & 0 & 1
       \end{matrix}
       \right].
\]

Now we change the graph by setting the length of edge $(7,\,9)$ as $2$, that is, $l_{79}=2$. When $N=3$ steps are allowed to transport a unit mass from node $1$ to node $9$, the evolution of mass distribution for the optimal transport plan, for $T=1$, is given by
 \[
       \tiny
       \left[
       \begin{matrix}
       1 & 0 & 0 & 0 & 0 & 0 & 0 & 0 & 0\\
       0 & 0.1554 & 0.4223 & 0.4223 & 0 & 0 & 0 & 0 & 0\\
       0 & 0 & 0 & 0 & 0 & 0 & 0.1554 & 0.8446 & 0\\
       0 & 0 & 0 & 0 & 0 & 0 & 0 & 0 & 1
       \end{matrix}
       \right].
\]
The mass travels through paths $1-2-7-9$, $1-3-8-9$ and $1-4-8-9$, but unlike the first case, the transport plan doesn't take equal probability for these three paths. Since the length of the edge $(7,\,9)$ is larger, the probability that the mass takes this path becomes smaller. The plan does, however, assign equal probability to the two paths $1-3-8-9$ and $1-4-8-9$ with minimum length, that is, these are the most probable paths. The evolutions of mass for $T=0.1$ and $T=100$ are 
\[
       \tiny
       \left[
       \begin{matrix}
       1 & 0 & 0 & 0 & 0 & 0 & 0 & 0 & 0\\
       0 & 0 & 1/2 & 1/2 & 0 & 0 & 0 & 0 & 0\\
       0 & 0 & 0 & 0 & 0 & 0 & 0 & 1 & 0\\
       0 & 0 & 0 & 0 & 0 & 0 & 0 & 0 & 1
       \end{matrix}
       \right]
\]
and
\[
       \tiny
       \left[
       \begin{matrix}
       1 & 0 & 0 & 0 & 0 & 0 & 0 & 0 & 0\\
       0 & 0.3311 & 0.3344 & 0.3344 & 0 & 0 & 0 & 0 & 0\\
       0 & 0 & 0 & 0 & 0 & 0 & 0.3311 & 0.6689 & 0\\
       0 & 0 & 0 & 0 & 0 & 0 & 0 & 0 & 1
       \end{matrix}
       \right]
\]
respectively. We observe that, when $T=0.1$ the flow assigns almost equal mass to the three available paths, while, when $T=100$, the flow concentrate on the most probable paths $1-3-8-9$ and $1-4-8-9$. This is clearly a consequence of Theorem \ref{temperature}.

\section{Conclusions and outlook}
In the present paper, we considered transportation over strongly connected, directed graphs. The development built on our earlier work \cite{chen2016networks}. More specifically, we introduced as measure of efficiency for a given transportation plan the average path length (cost) in e.g., \eqref{internal}, and as a measure of robustness the entropy \eqref{eq:entropy}. This allowed us to 
explore efficient-robust transport plans via solving corresponding optimization problems. Important insights gained in the present work include the results on certain invariances of the Schr\"odinger's bridges -- the ``iterated bridge'' invariance property and the invariance of the ``most probable path''. We explained their relevance for efficient-robust transport over networks. We also considered the dependence of the optimal transportation schedule on the temperature parameter following similar ideas from statistical physics.  In this, we highlighted the connection between the Schr\"odinger's bridge problem and OMT. Specifically, the solution of the Schr\"odinger bridge problem near-zero temperature is an approximation to the solution of the corresponding OMT problem. The relevance of the conceptual framework developed here for assessing robustness of real world networks (e.g., communication networks \cite{Jonck}, biological \cite{Alon,Sandhu}, and financial \cite{Sandhu1}) will be the subject of future work.

\spacingset{.97}

\end{document}